\documentclass[epj]{svjour}
\usepackage{graphics,graphicx}
\begin{document}
\title{Complex networks with scale-free nature and hierarchical modularity}
\author{Snehal M. Shekatkar \and G. Ambika
}                     
\institute{Indian Institute of Science Education and Research, Pune 411008, India}
\date{Received: date / Revised version: date}
%
\abstract{
Generative mechanisms which lead to empirically observed structure of networked systems from diverse fields like biology, technology and social sciences form a very important part of study of complex networks. The structure of many networked systems like biological cell, human society and World Wide Web markedly deviate from that of completely random networks indicating the presence of underlying processes. Often the main process involved in their evolution is the addition of links between existing nodes having a common neighbor. In this context we introduce an important property of the nodes, which we call mediating capacity, that is generic to many networks. This capacity decreases rapidly with increase in degree, making hubs weak mediators of the process. We show that this property of nodes provides an explanation for the simultaneous occurrence of the observed scale-free structure and hierarchical modularity in many networked systems. This also explains the high clustering and small-path length seen in real networks as well as non-zero degree-correlations. Our study also provides insight into the local process which ultimately leads to emergence of preferential attachment and hence is also important in understanding robustness and control of real networks as well as processes happening on real networks.
} 
\maketitle
\section{Introduction}
\label{intro}

Complex networks have become a key tool for understanding the behavior of complex systems from diverse fields like biology \cite{Jeong_2000,Jeong_2001,White_1986}, ecology \cite{Montoya_2002}, technology \cite{Albert_1999,Pastor_2001} and social sciences \cite{Newman_2001,Newman_2003}. Such networks are found to have interesting properties not possessed by their completely random counterparts and this indicates the presence of robust organizing principles behind their growth \cite{Networks_review}. The important features of such networks are existence of scale-free nature \cite{Barabasi_1999}, high clustering and small average path length \cite{Watts_1998}, hierarchical community structure or hierarchical modularity \cite{Ravasz_2002,Ravasz_2003,Clauset_2008,Lewis_2010} and non-zero degree correlations \cite{Newman_2002}. Over the years, various generative models of network growth have been proposed to explain one or more of these features \cite{Networks_review,Newman_book}. 

In scale-free network, the distribution of degrees follows a power law and this is usually attributed to preferential attachment \cite{Barabasi_1999}. The scale-free nature also implies small value for the average path length of the network \cite{Chen_2008}. Apart from being scale-free, most real networks are highly clustered as compared to completely random networks like Erd{\H o}s-R{\'e}nyi model \cite{Networks_review}. This means that in these networks, two nodes which share a neighbor have quite high chance of themselves being connected to each other. This tendency can be quantified using either the global clustering coefficient $C$ \cite{Newman_Strogatz_2001} or the Watts-Strogatz clustering coefficient $C_{WS}$ \cite{Watts_1998} of the network. Several variants of the basic preferential attachment model of Barab{\'a}si and Albert (BA model) have been proposed to explain this high value of clustering along with scale-free structure \cite{Holme_2002}. In addition to scale-free structure, real networks also show modular organization of clustering. In such networks, there exist groups of nodes such that the nodes in the same group are very densely connected to each other whereas these groups themselves are connected to each other relatively sparsely. Such groups are known as communities in the network and and the network is said to possess community structure or modular structure \cite{Newman_review_2012}. Existence of such modular structure implies high value of clustering coefficient $C$ while the reverse is not true.  Also, in many of these modular networks, modularity is hierarchical. This means that on a global scale, network is divided into communities of various sizes and each of those communities itself shows a modular structure inside it \cite{Ravasz_2002,Ravasz_2003,Clauset_2008,Lewis_2010}. 

It is understood that, in real networks, local processes are responsible for the emergence of overall global structure of the network \cite{Albert_2000}. The present understanding is that the preferential attachment should be regarded as an outcome of system-dependent local processes between the nodes, rather than a mechanism in itself \cite{Vazquez_2003,Evans_2005}. For example it has been suggested that gene duplication could be responsible for the scale-free structure of protein interaction networks \cite{Vazquez_2002}. A possible mechanism reported for the local interactions which could produce scale-free networks with high clustering, is triadic closure where new links are added between the neighbors of a particular node \cite{Vazquez_2003,Davidsen_2002,Holme_2002,Bianconi_2014}. The models based on this mechanism can also give rise to simple or hierarchical community structure \cite{Vazquez_2003,Bianconi_2014}. In these models of triadic closure only new nodes are considered as sources of triadic closure. However, in a very general sense, in real networks,  many new links are being created between existing nodes also. We feel the possibility of such ``internal links" also should form an important part of the growth of such networks. 

In the present paper we identify a fundamental property of hubs of complex networks that is almost system independent. We show that this property can potentially explain the simultaneous existence of scale-free structure and hierarchical modularity even with inclusion of internal links. Our model for growth of complex networks is also based on the concept of triadic closure but with an additional property for nodes called mediating capacity. This mediating capacity is a degree dependent property such that high degree nodes are weak mediators of triadic closure. As we argue in the following sections, hubs in any real network have capability to start and maintain interactions with diverse types of nodes. This makes the triadic closure around the hubs less likely and hence results in low value for their clustering coefficients endowing the network with hierarchical modularity.

The rest of the paper is organized as follows. Section \ref{Sec_Triadic_closure} describes the existence of triadic closure in various real world networks along with the social networks. In Section \ref{Sec_hubs_weak} we introduce the concept of mediating capacity. Section \ref{Sec_med_attach} describes our model for growth of complex networks called mediated attachment model. In Section \ref{Sec_simul}, we show the existence of hierarchical community structure in this model along with scale-free structure. Section \ref{Sec_characteristics} deals with time/size dependence of various important characteristics of the mediated network. Finally, in section \ref{Sec_conclude} we draw conclusions related to mediated attachment models and indicate future directions of study.

\section{\label{Sec_Triadic_closure} Existence of triadic closure in complex networks}
The triadic closure is proposed as one of the local processes that seem to be present in many real networks \cite{Holme_2002,Bianconi_2014}. This involves the formation of links between a pair of nodes of the networks which share one or more neighbors in the network (See Fig.~\ref{triadic_mediated}(a)). We now describe the existence of mechanisms analogous to triadic closure in various real networks. 

In the case of social networks like acquaintance, friendships, coauthors or Facebook networks, triadic closure is obviously present since usually people get to know each other through a common neighbor in the network. Languages are also viewed as a complex network of words which are connected to each other if they frequently occur in the same sentence \cite{Cancho_2001,Cancho_2004}. During the evolution of a particular language, when words $A$ and $B$ start appearing together in a sentence and if the same happens for words $B$ and $C$, with a fairly high probability words $A$ and $C$ would appear in the same sentence. In the case of web pages, most web pages keep evolving by the addition and modification of their content and hence new hyperlinks are usually added to the page. Suppose a web page A has hyperlink to another page B, which itself contains hyperlink to page C, then A can know about C through B and can connect to C. The co-citations also work the same way where instead of hyperlinks, bibliographies are used to know and connect to second neighbors \cite{Vazquez_2003}.

All the network growth models involving triadic closure models assume that no direct preferential attachment is present while network is growing and preferential attachment is only an outcome of local process of triadic closure \cite{Vazquez_2003,Evans_2005,Bianconi_2014}. Such triadic closure models have been proved to be extremely successful in producing highly clustered scale-free networks. However the models of triadic closure proposed so far usually consider only new node as the source of triadic closure \cite{Holme_2002,Vazquez_2003,Saramaki_2004,Bianconi_2014}. As noted earlier, in real networks majority of links are in fact created by old nodes. So we have to include the formation of such internal links in the growth model. This requires the introduction of a new property of hubs which makes them weak mediators such that the generated network are scale-free and hierarchically modular. 

\section{\label{Sec_hubs_weak}Hubs as weak mediators}

We now make an important observation about the nodes of all complex networks, especially scale-free networks. Usually hubs in scale-free networks are considered as most important nodes. We claim that the high degree nodes or hubs in complex networks have capability to interact with nodes of a wide range of properties. If we consider Google web page which is a hub in world wide web, the web pages which link to it are of diverse nature. Similarly, a hub in social network usually has interest in wide variety of topics and hence develops the connections to people from highly diverse fields. As our third example, we consider a network of words of a particular language. In this network, hub is usually a word that capable of appearing with words belonging to diverse range of contexts. For example, the word ``THE" in English can appear with many different words. In the jargon of network theory, we can thus say that hubs are usually connected to dissimilar nodes. On the other hand, a low degree node (a Wikipedia page about a technical topic from mathematics) would have links only to nodes similar to it and hence usually its neighbors are also similar with each other. 

In all these cases the probability that such diverse or dissimilar nodes will get connected is very low. This leads to the observation that hubs should function as weak mediators in such networks. For example, a hub in social network may introduce two of his contacts with each other but those two contacts would usually share different interests and hence the link may not get established between them. This implies that the mediating capacity of a node decreases with increase in degree. As we will see, this insight naturally unifies the emergence of hierarchical community structure with scale-free nature in such networks. 

\section{\label{Sec_med_attach}Mediated attachment model}

We present a generative mechanism in the growth of networks with the property of mediating capacity for the nodes as discussed in the previous section. To incorporate this, we conjecture that the average number of links mediated by a node with degree $k$ decreases with $k$. Therefore we take the probability that a node $i$ with degree $k_{i}$ connects two of its neighbors at a given time to be $1/k_{i}^{n}$  where $n$ is a parameter. For a node of degree $k$, maximum $^{k}C_{2}\sim k^{2}$ links are possible among its $k$ neighbors. Hence a node with degree $k$ will form approximately $k^{2}(1/k^{n}) = (1/k^{n-2})$ links at any given time. Thus we must have $n > 2$ if hubs are to behave as weak mediators. We also describe the effect of $n<2$ (i.e. the effect of hubs not being weak mediators) later in the section. 

Since real networks operate locally, usually what happens in one part of the network doesn't directly affect the processes like addition of links and nodes in the other part of the network. This means that both the number of links and number of nodes in the network that join the network is an increasing function of the size of the network. To incorporate this type of increase in the number of nodes, at each discrete time $t$ we add approximately $pN(t)$ number of nodes to the network where $p$ is the parameter with value between $0$ and $1$ and $N(t)$ is the number of nodes in the network at time $t$. The case of addition of one node at a time would then correspond to $p$ value that decreases with time. 

The growth of the network in this model thus involves two processes:
\begin{enumerate}
\item{Starting from two connected nodes at time $t = 0$, $\lceil{pN}\rceil$ number of nodes are added to the network at each discrete time $t$ where $\lceil{}\rceil$ denotes the ceiling function. Each new node connects to one of the old nodes randomly with uniform probability.}
\item{At the same time, every existing node $i$ of degree $k_{i}(t)$ $(>1)$ connects every pair of its neighbors with probability:
\begin{equation}
\label{med_prob}
w_{i}(t) = \frac{A}{k_{i}(t)^{n}}
\end{equation}
where $A$ and $n$ are adjustable parameters with $n>2$. The quantity $w_{i}(t)$ reflects the mediating capacity of node $i$ at time $t$.
}
\end{enumerate}

\begin{figure}
\resizebox{0.5\textwidth}{!}{%
  \includegraphics{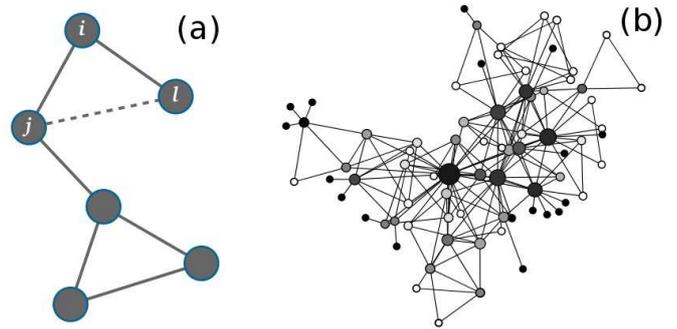}
}
\caption{ (a) A graphical representation of triadic closure process: node $i$ acts as a mediator for nodes $j$ and $l$ and tries to connect them. (b) Mediated network at time $t = 25$ with  $p = 0.1$, $A = 8$ and $n = 3$. The sizes of nodes are proportional to their degree values and the color of each node is graded according to its clustering coefficient, the darkest being the lowest.}
\label{triadic_mediated}
\end{figure}

A small network generated under these processes is shown in Fig.~\ref{triadic_mediated}(b). 

It can be easily seen that since each node is allowed to connect its neighbors with each other, the number of new links added to the network is an increasing function of the network size. If the rate of increase of network size is low, this would produce a highly dense network whose average degree $<k>$ would increase with time. However, in the present model, number of nodes also increases quite fast and as we will see below, these two factors can stabilize $<k>$ of the network.

\begin{figure}
\resizebox{0.5\textwidth}{!}{%
  \includegraphics{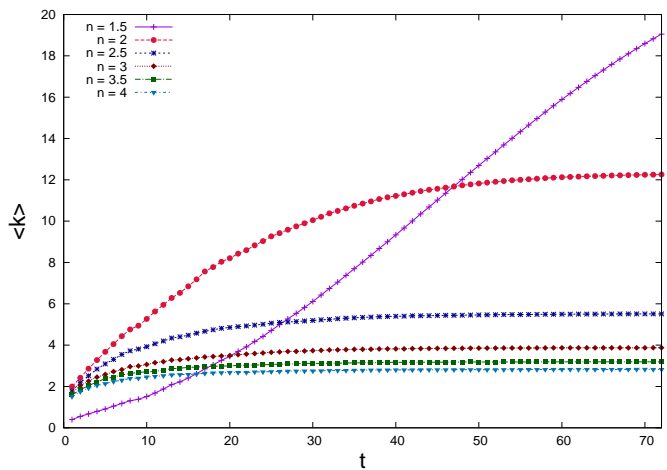}
}
\caption{ (Color online) Average degree of the network as a function of time for various values of $n$ with $A = 4$ and $p = 0.1$. For better visualization, for $n = 1.5$ the plotted values are equal to actual values divided by $5$. For $n < 2$, $<k>$ is a rapidly increasing function of time and hence leads to a non-stationary degree distribution. All cases with $n\ge 2$ lead to a stabilized averaged degree. In each case, network is grown up to size $N = 9525$ and results are averaged over $100$ realizations.}
\label{ave_deg_vs_time_A4}
\end{figure}

For the case of $n < 2$ we numerically compute the average degree $<k>$ as a function of time and find that it increases monotonically. This implies high link density and non-stationarity of the degree distribution. In general $<k>$ of the network must stabilize if the degree distribution is to be stationary asymptotically. In Fig.~\ref{ave_deg_vs_time_A4}, $<k>$ is plotted as a function of time for different values of $n$. The values for $n = 1.5$ are divided by $5$ for better visualization. Clearly, for $n < 2$,  $<k>$ is a very rapidly increasing function of time and hence the degree distribution is not stationary in this case whereas for $n>2$, it always stabilizes leading to a stationary degree distribution. 

\begin{figure}
\resizebox{0.5\textwidth}{!}{%
  \includegraphics{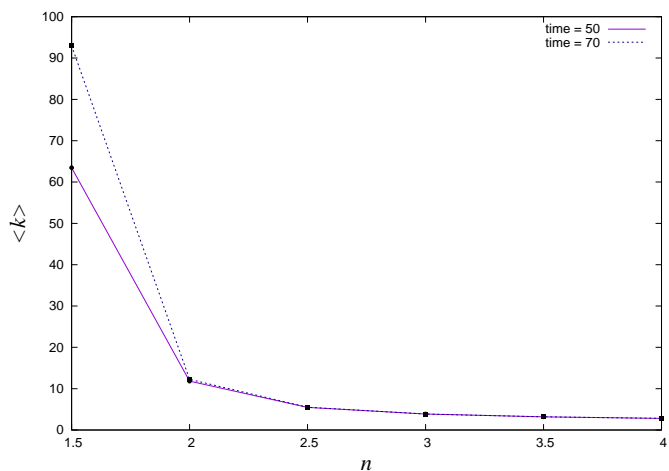}
}
\caption{(Color online) Average degree $<k>$ as a function of parameter $n$ at two different times $t = 50$ (a continuous line) and $t = 70$ (a solid line). It can be seen that when $n < 2$, $<k>$ has tendency to increase rapidly while for $n > 2$ average degree saturates. The results are averaged over $100$ random realizations.}
\label{deg_vs_n} 
\end{figure}

In Fig.~\ref{ave_deg_vs_time_A4} we show $<k>$ as a function of time for different values of $n$ for networks grown for $75$ units of time. With $p = 0.1$, the results shown are for $100$ realizations of the growth process. Fig.~\ref{deg_vs_n} shows the behavior of average degree as a function of parameter $n$ at two different times. The rapid increase in average degree with time (and hence the existence of non-stationary degree distribution) is evident. This behavior is understandable because for small $n$, the process of creation of links overtakes the process of creation of nodes. For the mediating capacity chosen as in Eq.(\ref{med_prob}) from $n = 2$ onwards these two effects balance each other stabilizing $<k>$.

\begin{center}
\begin{figure}
\resizebox{0.5\textwidth}{!}{%
  \includegraphics{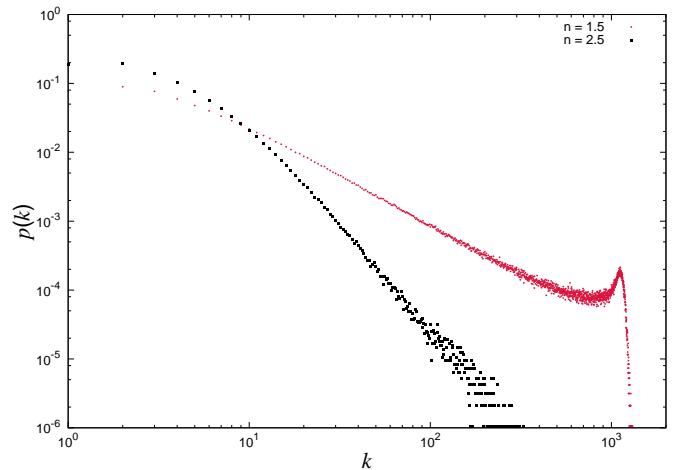}
}
\caption{ (Color online) Comparison of degree distributions for mediated network with $n<2$ and $n>2$. Here for $n < 2$ ($n = 1.5$ in this case), the distribution completely deviates from the power law in the tail and is also non-stationary as shown in Fig.~\ref{ave_deg_vs_time_A4}. Both cases correspond to $N = 9525$ and results are averaged over $100$ realizations.}
\label{compare_distri_A4}
\end{figure}
\end{center}

Also when $n < 2$, the network is so dense that there are too many nodes with very high degree and hence the network does not exhibit a scale-free structure. To see this, we plot the degree distributions of the network for $n = 1.5$ and $n = 2.5$ in Fig.~\ref{deg_distri_with_n_A8}. For $n = 1.5$, there exists a peak in the tail and hence the network in this case is not scale-free.

\section{\label{Sec_simul} Simultaneous occurrence of scale-free structure and hierarchical modularity}
\begin{center}
\begin{figure}
\resizebox{0.5\textwidth}{!}{%
  \includegraphics{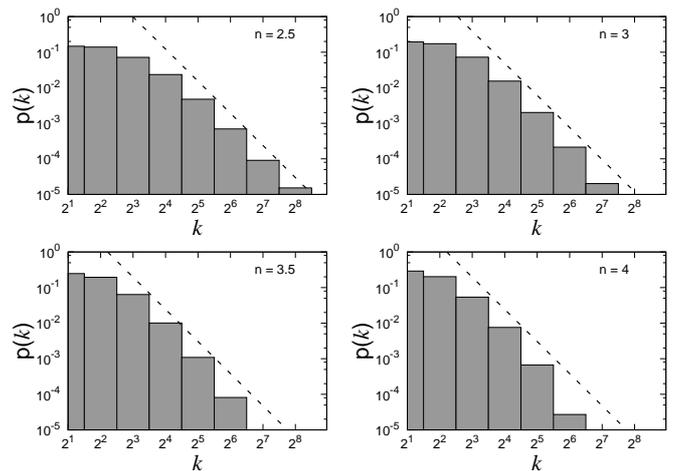}
}
\caption{  Degree distributions of mediated network of size $N = 9525$ for different values of $n$ with $A = 8$ and $p = 0.1$. A logarithmic binning is used to plot the histograms. The dotted line in each plot has slope $-3$ which is close to the actual scaling indices of real networks. All the results are averaged over $1000$ random realizations.}
\label{deg_distri_with_n_A8}
\end{figure}
\end{center}

Many real networks are both scale-free and hierarchically modular. The simultaneous occurrence of scale-free structure and hierarchical modularity is a less explored topic in the context of generative network models. As we now show, the mediated attachment model based on the conjecture that the hubs are weak mediators naturally gives rise to networks with both these properties. To establish the scale-free structure of the resulting network, in Fig.~\ref{deg_distri_with_n_A8} we plot the degree distributions of the network for various values of $n$ with $n > 2$ and all of them can be seen to follow a power-law behavior with a scaling index close to $3$. The real networks are usually seen to have scaling indices in the range $(2,4)$ \cite{Networks_review,Newman_book} and hence the model can be tuned to produce the actually observed distributions satisfactorily. 

\begin{figure}
\resizebox{0.5\textwidth}{!}{%
  \includegraphics{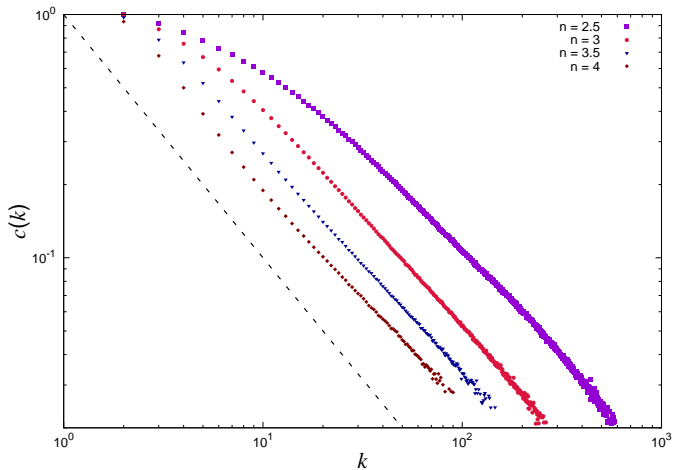}
}
\caption{(Color online) Local clustering coefficient $c$ as a function of degree $k$ for mediated network. The parameters are same as for Fig.~\ref{deg_distri_with_n_A8}. The black dotted line has slope $-1$. Results are averaged over $1000$ random realizations.}
\label{cluster_vs_deg_with_n_A8}
\end{figure}

A necessary (but not sufficient) indicator of existence of hierarchical modularity in the given network is the dependence of local clustering coefficient on the degree. A local clustering of the node $i$ is given as:

\begin{equation}
c_{i} = \frac{E_{i}}{^{k}C_{2}}
\end{equation}
where $k$ is the degree of node and $E_{i}$ is the number of links present among the neighbors of node $i$. Its value always lies between $0$ and $1$. For the networks with hierarchical modularity, $c_{i}$ shows a systematic decrease with degree $k$ with $c(k)\sim k^{-1}$ \cite{Ravasz_2003}. For the model presented here, we compute the local clustering coefficient for different values of $n$ and plot it as a function of $k$ in Fig.~\ref{cluster_vs_deg_with_n_A8}. The decrease of $c(k)$ with degree is clear in all cases. This happens because the mediating capacity of a node decreases with increase in its degree and which in turn reduces its clustering coefficient when degree is large.

To confirm that the mediated network does possess the hierarchical modularity, we use hierarchical Infomap (http://www.mapequation.org/code.html). We find that for all $n > 2$, the mediated network is hierarchically modular but the number of modular levels depends on the value of $n$ and in fact is the increasing function of $n$. Thus, for the network size $N = 9525$ used here, for $n = 2.5$ network has $4$ to $5$ levels of hierarchy while for $n = 4$, the network has $8$ to $9$ levels of hierarchy. 

\section{\label{Sec_characteristics} Characteristics measures and their scaling}
Now we look at the various important network characteristics like clustering coefficient, path length and degree correlations for the mediated network.
\subsection{Clustering coefficient}
The density of triangles in the network is quantified by the quantity called clustering coefficient. There exist two definitions of the clustering coefficient in the literature: global clustering coefficient and Watts-Strogatz clustering coefficient. In the present case, we use Watts-Strogatz clustering coefficient $C_{WS}$ (also called as average clustering coefficient) as the quantifier for the clustering in the network. It is defined as the average of the local clustering coefficients over all the nodes of the network:

\begin{equation}
C_{\tiny{WS}} = \frac{1}{N}\sum\limits_{i=1}^{N}c_{i}
\end{equation}

\begin{figure}
\resizebox{0.5\textwidth}{!}{%
  \includegraphics{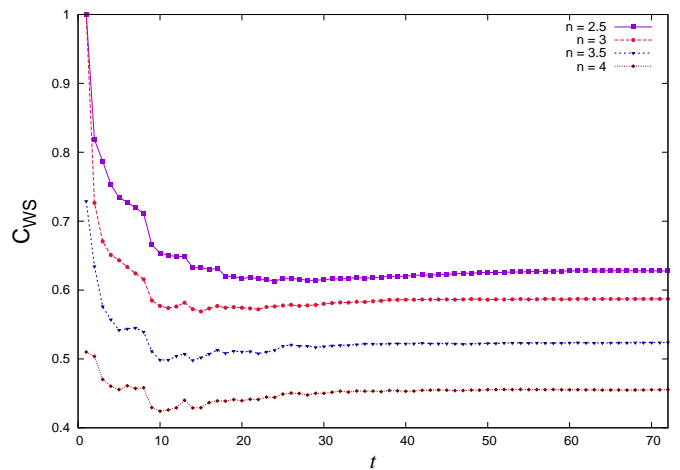}
}
\caption{ (Color online) Watts-Strogatz clustering coefficient $C_{WS}$ of the network as a function of time for various values of parameter $n$. All the parameter values are same as in Fig.~\ref{cluster_vs_deg_with_n_A8}}
\label{WS_cluster_with_n_A8}
\end{figure}

Fig.~\ref{WS_cluster_with_n_A8} shows the Watts-Strogatz clustering coefficient of the mediated network as a function of time for various values of $n$. It can be seen that all $C_{WS}$ values settle asymptotically to values in the range $(0.4,0.7)$. Almost all the real networks like film actors, coauthorships, citations, Internet, WWW, metabolic networks, protein interactions, ecological food webs and neural networks show such high values of $C_{WS}$ \cite{Newman_review}. This agreement with the empirical observations further supports the model presented here.

\subsection{Degree correlations}

Apart from the properties discussed so far, many real networks show non-zero degree correlations. Networks in which high degree nodes tend to connect to low degree nodes (i.e. networks for which negative degree correlations exist) are known as dissortative networks while those in which similar degree nodes tend to connect to each other (i.e. networks for which positive degree correlations exist) are known as assortative networks. 

Assortative or dissortative nature of networks can be quantified by calculating the assortativity coefficient for the network which is given by the following expression \cite{Newman_book}:

\begin{equation}
r = \frac{\sum\limits_{ij}(B_{ij}-k_{i}k_{j}/2m)k_{i}k_{j}}{\sum\limits_{ij}(k_{i}\delta_{ij}-k_{i}k_{j}/2m)k_{i}k_{j}}
\end{equation}

Here, $B_{ij}$ is $(i,j)^{th}$ element of an adjacency matrix of the network, $m$ is the number of links in the network, $k_{i}$ is the degree of node $i$ and $\delta_{ij}$ is the Kronecker delta whose value is $1$ when $i = j$ and is $0$ otherwise.

\begin{figure}
\resizebox{0.5\textwidth}{!}{%
  \includegraphics{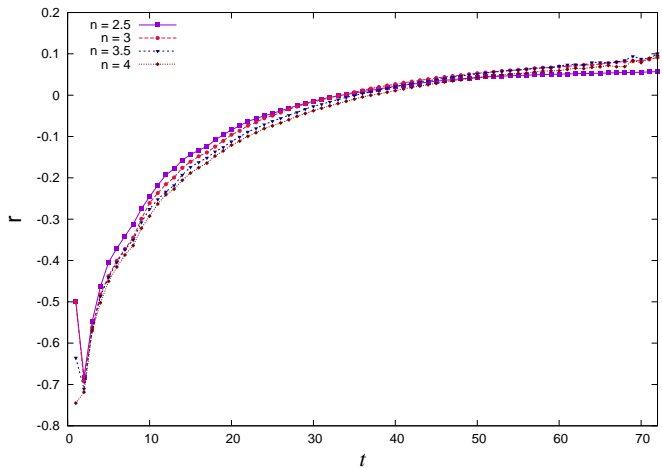}
}
\caption{ (Color online) Assortativity coefficient $r$ of the mediated network as a function of time for different values of parameter $n$. It can be seen that $r$ stabilizes to positive values asymptotically and hence the mediated network is assortative. All the parameter values are as in Fig.~\ref{cluster_vs_deg_with_n_A8}}
\label{assort_with_n_A8}
\end{figure}

Interestingly, social networks are always seen to possess positive degree correlations i.e they are assortative \cite{Newman_2002}. In Fig.~\ref{assort_with_n_A8} we show the assortativity coefficient of the mediated network as a function of time for different $n$. It is clear that our model with $n > 2$ leads to positive degree correlations. 

\subsection{Average path length}
\begin{figure}
\resizebox{0.5\textwidth}{!}{%
  \includegraphics{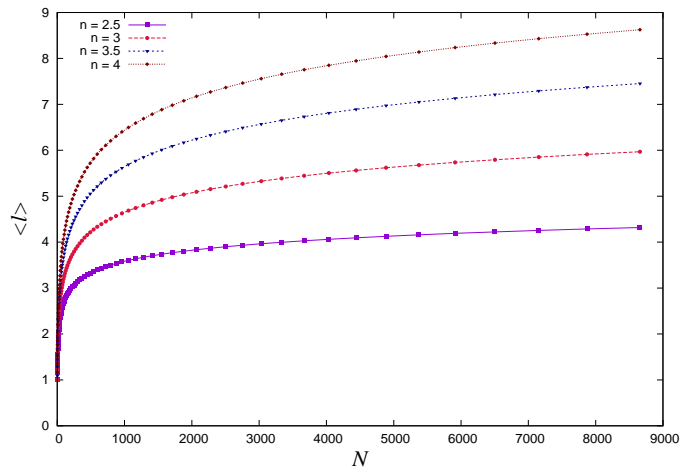}
}
\caption{(Color online) Average path length $<l>$ of the network as a function of size $N(t)$ of the network for different $n$. The parameter values are as given in Fig.~\ref{cluster_vs_deg_with_n_A8}. Each curve in the plot follows a logarithmic increase.}
\label{path_with_n_A8}
\end{figure}

The real networks are found to have a small-world property which means that the average path length $<l>$ of these networks varies very slowly with size and also they have high clustering coefficient \cite{Watts_1998}. We also calculate the $<l>$ of mediated network as a function of its size for different $n$. It is clear that $<l>$ increases logarithmically with size (see Fig.~\ref{path_with_n_A8}). Such logarithmic scaling with size has been reported earlier for many real networks \cite{Networks_review}. This along with the observed high clustering, makes the mediated network a small-world network. 

\section{\label{Sec_conclude}Conclusion}
Understanding the physical mechanisms which lead to the emergence of various structural properties seen in real networks is an important part of the network science. Since the discovery of small-world nature and scale-free nature of various real networks, better and better models of network formation have been proposed over the years. It has been observed that many real networks exhibit both scale-free nature and hierarchical modularity. However, the generic mechanisms which lead to their simultaneous occurrence in real networks are not yet very clear. In the present work we showed that the weakness of hubs in mediating the connections between their neighbors produces both of these structures successfully. The weakness of hubs can be understood from a more fundamental fact that hubs are usually connected to dissimilar nodes in the network. We feel that a more careful use of this idea in the context of specific networks will lead to many novel insights into the structure and function of various real networks. 

The model presented here comes with tunable parameters $p,A$ and $n$ that can be adjusted to get desired values of network characteristics. For $n>2$ we showed that the average degree of the network stabilizes and can be tuned to a desired value by tuning parameters $A$ and $n$. Apart from scale-free nature and hierarchical modularity, we also showed that the mediated network is highly clustered and its path length scales logarithmically with size. This shows that a small-world nature is also an emerging property of this model. Finally, the network shows positive degree correlations as observed in case of social networks. We anticipate that detailed investigations to understand the intrinsic mechanisms that generate this property in different cases, can reveal interesting processes underlying the complexity of such systems.

\vspace{10pt}
One of the authors (S.M.S.) would like to thank University Grants Commission, New Delhi, India for financial assistance in the form of Senior Research Fellowship. 
\providecommand{\noopsort}[1]{}\providecommand{\singleletter}[1]{#1}%

\end{document}